\documentclass[journal,draftclsnofoot,onecolumn,12pt]{IEEEtran}
\usepackage{amsmath,amsfonts}
\usepackage{algorithmic}
\usepackage{algorithm}
\usepackage{array}
\usepackage[caption=false,font=normalsize,labelfont=sf,textfont=sf]{subfig}
\usepackage{textcomp}
\usepackage{stfloats}
\usepackage{url}
\usepackage{verbatim}
\usepackage{graphicx}
\usepackage{cite}
\usepackage{color}
\usepackage{amsthm}
\usepackage{balance}
\usepackage{epstopdf}
\usepackage{latexsym,bm}
\newtheorem{theorem}{Theorem}

\newtheorem{corollary}{Corollary}

\makeatletter
\newlength{\figwidth}
\ifCLASSOPTIONonecolumn
\else
\fi
\begin{document}
\title{On the Design of Capacity-Achieving Distributions for Discrete-Time Poisson Channel with Low-Precision ADCs}

\author{Qianqian Li,
Lintao Li,
Lixiang Liu,
Lei Yang,
Caihong Gong,
Hua Li,
Shiya Hao, and
Xiaoming Dai %,~\IEEEmembership{Staff,~IEEE,}
        % <-this % stops a space
%\thanks{This paper was produced by the IEEE Publication Technology Group. They are in Piscataway, NJ.}% <-this % stops a space
%\thanks{Manuscript received April 19, 2021; revised August 16, 2021.}
%\thanks{Copyright (c) 20xx IEEE. Personal use of this material is permitted. However, permission to use this material for any other purposes must be obtained from the IEEE by sending a request to pubs-permissions@ieee.org.}
%\thanks{This work was supported in part by the National Natural Science Foundation of China under Grant 62371037, in part by the Guangdong Basic and Applied Basic Research Foundation under Grant 2024A1515011186 and Grant 2025A1515010082, in part by Project "Vice President of Science and Technology" of Changping District, Beijing under Grant 202304001017, in part by the China Postdoctoral Science Foundation under Grant 2023M740223, and in part by the National Natural Science Foundation of China under Grant U24A20214. (\emph{Corresponding author: Xiaoming Dai})}
\thanks{ %Institute of Software, Chinese Academy of Sciences, Beijing 100080, P.R China
Q.~Li and X.~Dai are with the School of Computer and Communication Engineering, University of Science and Technology Beijing, Beijing 100083, China, and also with the Shunde Innovation School, University of Science and Technology Beijing, Foshan 528399, China (email: liqianqian\_lf@163.com, daixiaoming@ustb.edu.cn).\\
{Corresponding author: Xiaoming Dai}.}
}
\markboth{Journal of \LaTeX\ Class Files,~Vol.~14, No.~8, August~2021}%
{Shell \MakeLowercase{\textit{et al.}}: A Sample Article Using IEEEtran.cls for IEEE Journals}
\maketitle
\begin{abstract}
%This paper investigates the design of capacity-achieving distribution for the discrete-time Poisson channel (DTPC) with low-precision analog-to-digital converters (ADCs) under dark current effects.
This paper investigates the design of the capacity-achieving input distribution for the discrete-time Poisson channel (DTPC) under dark current effects with low-precision analog-to-digital converters (ADCs). 
This study introduces an efficient optimization algorithm that integrates the Newton-Raphson and Blahut-Arimoto (BA) methods to determine the capacity-achieving input distribution and the corresponding amplitudes of input mass points for the DTPC, subject to both peak and average power constraints.
%We propose an efficient optimization algorithm integrating the Newton-Raphson and the Blahut-Arimoto (BA) methods to derive the capacity-achieving distribution and the amplitude of the input mass points for the quantized DTPC under peak power and average power constraints.
 Additionally, the Karush-Kuhn-Tucker (KKT) conditions are established to provide necessary and sufficient conditions for the optimality of the obtained capacity-achieving distribution.
Simulation results illustrate that the proposed algorithm attains $72\%$ and $83\%$ of the theoretical capacity at 5 dB for 1-bit and 2-bit quantized DTPC, respectively. Furthermore, for a finite-precision quantized DTPC (i.e., ${\log _2}K$ bits), the capacity can be achieved by a non-uniform discrete input distribution with support for $K$ mass points, under the given power constraints.
\end{abstract}


\section{Introduction}
\IEEEPARstart{T}{he} discrete-time Poisson channel (DTPC) is a well-studied model of direct detection optical communication, where the transmitter modulates the intensity of the optical signal for transmission, and the receiver performs direct detection using photodetectors \cite{paper_11}. 
{{ Compared to radio-frequency (RF) systems, laser links provide higher bandwidth and lower latency, which are critical for time-sensitive data transmission contexts (e.g., disaster monitoring). 
%The DTPC model is widely adopted in these scenarios as it accurately characterizes photon-counting statistics in optical receivers.
For example, low earth orbit satellites (e.g., earth observation or military satellites) generate massive data (e.g., high-resolution images, real-time surveillance) that must be downlinked within brief overflight windows (typically 5–10 minutes). 
%Failure to transmit within this window risks data loss due to orbital motion 
Failure to complete the transmission within this constrained period may lead to data loss due to the dynamics of orbital motion \cite{paper_002}. }}

{{The objective of input distribution design is to optimize how the information is transmitted over the given channel, thereby maximizing the achievable data rate and enabling the system to approach its capacity limits.}} Recent research on DTPC has primarily focused on identifying optimal capacity-achieving distributions \cite{paper_15aa,paper_13,paper_15,paper_14}. 
In \cite{paper_15aa}, the authors studied the capacity-achieving distribution of the DTPC with peak power constraint, concluding that the distribution is unique and discrete with finite support. In \cite{paper_13}, a deterministic annealing (DA) algorithm was investigated to obtain the capacity and capacity-achieving distributions for the DTPC with peak power and average power constraints. The authors in \cite{paper_15} constructed a duality-based framework to derive an improved capacity upper bound, which illustrated that the capacity-achieving distribution follows a discrete distribution for the channel without dark current and with an average power constraint.
Additionally, the authors in \cite{paper_14} investigated the asymptotic capacity at low input powers for the DTPC with an average power constraint or both peak power and average power constraints.
In order to get more insights about the method for capacity-achieving distribution. In \cite{paper_24}, the Blahut-Arimoto (BA) algorithm was proposed to compute the capacity of an arbitrary discrete memoryless channel, though it did not account for the amplitude of the input mass points. Matz et al. \cite{paper_25} presented accelerated BA and natural gradient algorithms that converge significantly faster than the traditional BA algorithm. In \cite{paper_27}, the cutting-plane algorithm was utilized to calculate capacity and optimize the input distribution for discrete-time memoryless Additive White Gaussian Noise (AWGN) quantization channels. However, this algorithm exhibits high computational complexity when applied to multi-bit analog-to-digital converters (ADCs) quantization scenarios.
The aforementioned research works provide valuable insights and methods for obtaining the capacity and capacity-achieving distribution of the DTPC. However, the capacity is approaching the Shannon limit with the growth of wireless communication rates.
Therefore, exploring methods to address the high complexity and power consumption of receivers while simultaneously increasing channel capacity of the DTPC is crucial.

{While traditional studies on DTPC focus on full-precision ADCs, these high-precision ADCs are often impractical for energy-constrained systems such as satellites, unmanned aerial vehicles (UAVs), and portable optical transceivers. %High-precision ADCs consume significant power consumption and manufacturing costs, and increase system complexity,  making them unsuitable for compact, energy-efficient payloads 
High-precision ADCs entail substantial power consumption and manufacturing costs, while also increasing system complexity, rendering them impractical for compact, energy-efficient payloads \cite{paper_19}. 
In contrast, low-precision ADCs offer a substantial reduction in power consumption, lower data storage and transmission demands, and enable the development of high-speed communication systems with manageable energy consumption, which has garnered significant attention in the academic community \cite{paper_17}. As a result, modern communication systems increasingly adopt low-precision ADCs as they strike a favorable balance between power efficiency and performance \cite{paper_18} \cite{paper_21}.}
The authors in \cite{paper_20} analyzed the spectral efficiency of single-carrier and orthogonal frequency division multiplexing (OFDM) transmission in massive multiple-input multiple-output (M-MIMO) systems with one-bit ADCs. An additive quantization noise model (AQNM) was considered to describe the quantization distortion in a cell-free M-MIMO system, and the closed-form expression of the system rate was derived in \cite{paper_22}.
In \cite{paper_23}, channel estimation for uplink multi-user M-MIMO systems using one-bit ADCs to quantize the received signal was investigated. These studies mentioned above have demonstrated that low-precision ADCs can significantly reduce hardware costs and power consumption in M-MIMO systems, providing a solid foundation and comprehensive understanding of low-precision ADCs.

{In addition, dark current, an inherent noise source in photodetectors, significantly impacts system performance, particularly in low-light and weak-signal scenarios.
This work on the low-precision quantized DTPC model with dark current bridges the critical gap between theory and practical implementation, thereby enhancing its relevance and applicability to real-world systems.} %Although some general characteristics of capacity-achieving distributions for the DTPC are known, those of the DTPC with low-precision ADC and the corresponding analytical framework are remain understudied. 
While certain general properties of capacity-achieving distributions for the DTPC are well-established, those specific to the DTPC with low-precision ADCs, along with the associated analytical framework, remain underexplored.
Consequently, we propose an alternating optimization (AO) algorithm, leveraging the Newton-Raphson and BA methods, to address the capacity maximization problem for the quantized DTPC.
%Therefore, we propose an alternate optimization (AO) algorithm based on the Newton-Raphson and the BA methods approach to the capacity maximization problem of the quantized DTPC. 
The peak power and average power constraints are incorporated into the channel input, and the capacity-achieving distribution along with the input mass point's amplitude are obtained using the Armijo rule and a two-symbols-one-location strategy. In the proposed algorithm, a Newton-Raphson method replaces the interval-halving procedure of the BA algorithm in \cite{paper_17aa}.
The output of the proposed algorithm can be tested for optimality against the Karush-Kuhn-Tucker (KKT) conditions. Additionally, this paper analyses the impact of quantization on the capacity of the DTPC. Developing numerical methods to compute channel capacity and optimal input distribution for the quantized DTPC is of practical importance, as these numerical results are beneficial for system implementation.

The rest of this paper is organized as follows. In Section II, we construct a low-precision ADCs quantized DTPC model. In section III, an AO algorithm based on the Newton-Raphson and the BA methods is developed to obtain capacity-achieving input distributions for the low-precision ADCs quantized DTPC with peak power and average power constraints. In Section IV, we present the numerical results and discuss the impact of the proposed algorithm on capacity. Section V provides a summary of this work.

\emph{Notation}: In this paper, the main notations are shown as follows. $Q{\left(  \cdot  \right)}$ denotes the low-bit ADC quantizer. $\sum {\left(  \cdot  \right)}$ represents the operation of summation. We use $I\left( {A;B} \right)$ to denote the mutual information of $A$ and $B$. $\delta \left(  \cdot  \right)$ denotes the Dirac impulse function. $\frac{{\partial {f_X}\left( x \right)}}{{\partial x}}$ represents the partial derivative of ${f_X}\left( x \right)$ with respect to the independent variable $x$.

\section{System Model}
\begin{figure}[h]
\centering
\includegraphics[width=0.45\textwidth]{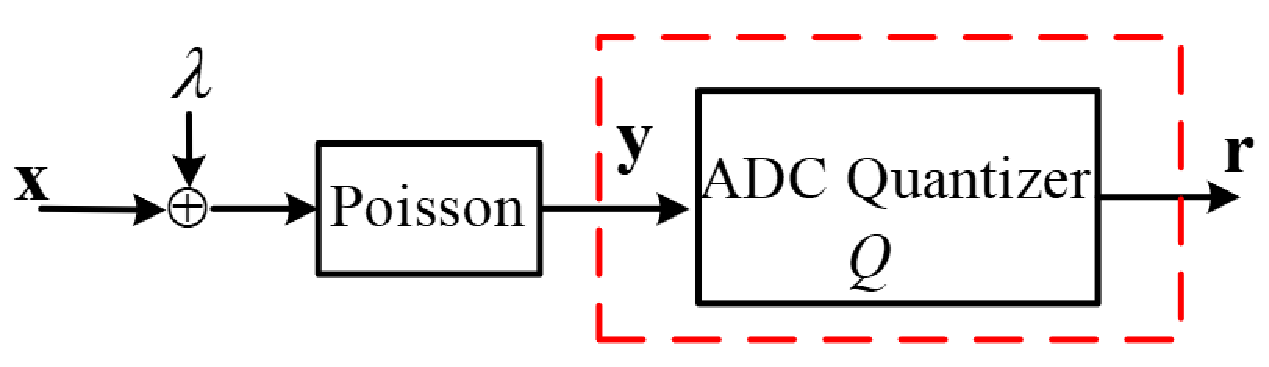}
\caption{Quantized DTPC model.}
\label{fig_1}
\end{figure}
We consider a well-studied model of direct detection optical communication systems restricted to piecewise constant pulse-amplitude modulation (PAM) transmission techniques. The quantized DTPC model is described in Fig.~\ref{fig_1}.
The input set consists of $N$ discrete transmitted symbols ${\bf{x}} = \left\{ {{x_1},{x_2},...,{x_N}} \right\}$ with ${x_i} \in \mathbb{R}_{0}^{+}$, $i \in \left( {1,2,...,N} \right)$.
 The channel output ${\bf{y}} = \left\{ {{y_1},{y_2},...,{y_N}} \right\}$, ${y_i} \in \mathbb{Z}_{0}^{+}$ denotes the number of photons that are detected at the receiver during the pulse duration. For ease of presentation, the symbol index subscript $i$ is omitted whenever possible. Conditioned on the input ${x \ge 0}$, the channel output ${y}$ is Poisson distributed with mean $x + \lambda $.  %We consider a quantized DTPC model as show in Fig. 1.
The channel law is given by
{\begin{align}
\label{equation_1}
 P_{\mathrm{y} \mid \mathbf{x}}(y \mid x)=\operatorname{Pois}_{x+\lambda}(y)=\frac{{{{(x + \lambda )}^y}}}{{y!}}\exp \left( { - (x + \lambda )} \right),
%x \in \mathbb{R}_{0}^{+}, y \in \mathbb{Z}_{0}^{+},
\end{align}}
where $\lambda  \ge 0$ is the dark current, representing the non-ideality of the detector and degrades the received counts even in the absence of illumination. $\lambda$ is a constant independent of the input $x$.

It is reasonable to impose certain constraints on the input of the DTPC in practical applications. 
The peak power and average power constraints are given by
\begin{equation}
\label{equation_2}
{\rm E}\left( {x} \right) \le \epsilon ,%\  {\rm{[Average\  Power\  Constraint]}}{\rm{,}}
\end{equation}
\begin{equation}
\label{equation_3}
0 \le {x} \le A,%\  {\rm{[Peak\  Power\  Constraint]}}{\rm{.}}
\end{equation}
respectively, where $0 \le \epsilon  \le A$.
The peak-to-average power ratio is defined by
\begin{equation}
\label{equation_4}
\alpha  = \frac{A}{\epsilon},
\end{equation}
where $\alpha  = 1$ implies the absence of the average power constraint, and ${1 \mathord{\left/
 {\vphantom {1 \alpha }} \right.
 \kern-\nulldelimiterspace} \alpha } \ll 1$ implies a weak peak power constraint.

 The receiver uses a thresholding quantizer $Q$, which is equivalent to a threshold vector ${\bf{q}} = \left[ {{q_1},{q_2},...,{q_{K- 1}}} \right]$ with ${{q_i}\in {\mathbb{R}_{0}}}$, $i \in \left( {1,...,K - 1} \right)$, and $0: = {q_0} < {q_1} < {q_2} < ... < {q_{K - 1}} < {q_K}: = \infty$.
 The output of the quantized DTPC is given by
 \begin{equation}
\label{equation_5}
{r_i} = Q\left( y \right),{\ }{\rm{ if }}{\ }{q_{j - 1}} < y \le {q_j}.
\end{equation}
Thus, the resulting channel transition probability function is given by
{\begin{align}
\label{equation_6}\nonumber
&{g_{{\bf{r}}\left| {\bf{x}} \right.}}\left( {r\left| x \right.} \right) = P\left( {{\bf{r}} = {r_j}\left| {{\bf{x}} = x} \right.} \right)\\
& = \sum\limits_{{y} = {q_{j - 1}}}^{{q_j}} {\frac{{{{\left( {x + \lambda } \right)}^{{y}}}}}{{{y}!}}{\exp \left( { - (x + \lambda )} \right)}} ,\ {\rm{        }}\ 1 \le j \le K{\rm{                 }}.
\end{align}}
The probability mass function (PMF) of ${\bf{r}}$ for the quantized DTPC is given by
\begin{equation}
\label{equation_7}
{g_{\bf{r}}}\left( r \right) = P\left( {{\bf{r}} = {r_j}} \right) = \sum\nolimits_x {{g_{{\bf{r}}\left| {\bf{x}} \right.}}\left( {r\left| x \right.} \right){p_{\bf{x}}}\left( x \right)},
\end{equation}
where ${p_{\bf{x}}}\left( x \right) = \{ {p\left( {{x_1}} \right),p\left( {{x_2}} \right),...,p\left( {{x_N}} \right)} \}$ is the input distribution.
The cumulative distribution function (CDF) of the input is denoted by ${F_x}$, which is written as
\begin{equation}
\label{equation_8}
d{F_x} = \sum\nolimits_{\bf{x}}{{p_{\bf{x}}}\left( {{x_i}} \right)\delta \left( {x - {x_i}} \right)},
\end{equation}
where $\sum\nolimits_{\bf{x}} {{p_{\bf{x}}}\left( x \right)}  = 1$.
{\section{THE PROPOSED ALGORITHM}
In this section, we propose an AO algorithm for the quantized DTPC model with dark current under peak power and average power constraints to derive the capacity-achieving input distribution.}
%This section proposes an AO algorithm approach to the quantized DTPC capacity maximization problem in (\ref{equation_9}) with peak and average power constraints
The capacity (in nats per channel use) of the quantized DTPC under the peak power and average power constraint is given by
\begin{align}
\nonumber
C &\buildrel \Delta \over  = \mathop {\max }\limits_{{F_x} \in f} I\left( {{\rm{\bf{x};\bf{r}}}} \right)\\\nonumber
&= \mathop {\max }\limits_{{F_x} \in f} \sum\nolimits_{\bf{x}} {\sum\nolimits_{\bf{r}}{{p_{\bf{x}}}\left( x \right){g_{{\bf{r}}\left| {\bf{x}} \right.}}\left( {r\left| x \right.} \right)\log \frac{{{g_{{\bf{r}}\left| {\bf{x}} \right.}}\left( {r\left| x \right.} \right)}}{{g_{\bf{r}}}\left( r \right)}} } \\\label{equation_9}
&= \mathop {\max }\limits_{{F_x} \in f} \sum\nolimits_{\bf{x}} {\sum\nolimits_{\bf{r}} {{p_{\bf{x}}}\left( x \right){g_{{\bf{r}}\left| {\bf{x}} \right.}}\left( {r\left| x \right.} \right)\log \frac{{{P_{{\bf{x}}\left| {\bf{r}} \right.}}\left( {x\left| r \right.} \right)}}{{{p_{\bf{x}}}\left( x \right)}}} },
\end{align}
where $f$ represents the set of all distributions that satisfy the peak power and average power constraints, which is expressed as
\begin{equation}
\label{equation_10}
f \buildrel \Delta \over = \!\left\{ \!{{F_x}\!:\!\sum\nolimits_{\bf{x}} {{p_{\bf{x}}}\left( x \right)} \! = \!1,\!\sum\nolimits_{\bf{x}} \!{x{p_{\bf{x}}}\!\left( x \right)} \! \le \!\epsilon ,0 \!\le\! x\! \le\! A} \!\right\}.
\end{equation}
\begin{theorem}\label{Theorem_1}
[KKT conditions \cite{paper_15aa}]%, Eq. (24),(25)
The capacity-achieving distribution is ${F_x^*}\left( {A, \epsilon} \right)$
if the following expressions are satisfied for some $\mu  \ge 0$
\begin{equation}
\label{equation_10a}
i\left( {x;{F_x^*}\left( {A, \epsilon} \right)} \right) \le I\left( {F_x^*} \right) + \mu \left( {x - \epsilon } \right),\forall x \in \left[ {0,A} \right],
\end{equation}
\begin{equation}
\label{equation_10b}
i\!\left( {x;{F_x^*}\left( {A, \epsilon} \right)} \right)\! =\! I\left( {F_x^*} \right)\! +\! \mu \left( {x\! -\! \epsilon} \right),\forall x \!\in\! \psi \left( {F_x^*\left( {A,\epsilon} \right)} \right),
\end{equation}
where $i\left( {x;{F_x^*}\left( {A, \epsilon} \right)} \right) =  - \sum\limits_{r = 0}^\infty  {{P_{{\bf{r}}\left| {\bf{x}} \right.}}\left( {r\left| x \right.} \right)} \log \frac{{{P_{\bf{r}}}\left( r \right)}}{{{P_{{\bf{r}}\left| {\bf{x}} \right.}}\left( {r\left| x \right.} \right)}}$ and $I\left( {{F_x^*}} \right) = \int_0^A {i\left( {x,{F_x^*}\left( {A, \epsilon} \right)} \right)} d{F_x^*}\left( {A, \epsilon} \right)$. $\psi \left( {F_x^*\left( {A,\epsilon} \right)} \right)$ denotes the set of increase points of ${F_x^*\left( {A,\epsilon} \right)}$.
This theorem states the Kuhn-Tucker conditions and the proof is classical \cite{paper_17bb}.
If (\ref{equation_10a}) and (\ref{equation_10b}) are not satisfied, another distribution $F_x^*\left( {A, \epsilon} \right) \ne {F_x}\left( {A, \epsilon} \right)$ can be constructed in the same way, which satisfies $i\left( {{F_x}} \right) > i\left( {F_x^*} \right)$. The KKT conditions provide sufficient and necessary conditions for the optimality of the capacity-achieving distribution.
\end{theorem}

For a given quantizer $Q$ and input distribution ${{p_{\bf{x}}}\left( x \right)}$, ${P_{{\bf{x}}\left| {\bf{r}} \right.}}\left( {x\left| r \right.} \right)$ is defined according to the Bayes theorem as
\begin{align}
\label{equation_11}
{P_{{\bf{x}}\left| {\bf{r}} \right.}}\left( {x\left| r \right.} \right) = \frac{{{p_{\bf{x}}}\left( x \right){g_{{\bf{r}}\left| {\bf{x}} \right.}}\left( {r\left| x \right.} \right)}}{{\sum\nolimits_{\bf{x}} {{p_{\bf{x}}}\left( x \right){g_{{\bf{r}}\left| {\bf{x}} \right.}}\left( {r\left| x \right.} \right)} }}.
\end{align}
Define $W$ as the Lagrangian associated with the optimization
problem as 
%the Lagrangian function is constructed as
\begin{align}
 \nonumber \label{equation_14}
W &\buildrel \Delta \over = \sum\nolimits_{\bf{x}} {\sum\nolimits_{\bf{r}} {{p_{\bf{x}}}\left( x \right){g_{{\bf{r}}\left| {\bf{x}} \right.}}\left( {r\left| x \right.} \right)\log \frac{{{P_{{\bf{x}}\left| {\bf{r}} \right.}}\left( {x\left| r \right.} \right)}}{{{p_{\bf{x}}}\left( x \right)}}} } \\
 &+\! {\lambda _0}\!\left(\! {\sum\nolimits_{\bf{x}} {{p_{\bf{x}}}\left( x \right)} \! -\! 1}\! \right)\! + \!\eta \left( \!{\sum\nolimits_{\bf{x}} {x{p_{\bf{x}}}\left( x \right)} \! -\! \epsilon} \!\right),
\end{align}
where ${\lambda _0} \ge 0$ and $\eta  \ge 0$ are the Lagrange multipliers.
\begin{corollary}\label{theorem_1}
For a given quantizer and conditional distribution ${P_{{\bf{x}}\left| {\bf{r}} \right.}}\left( {x\left| r \right.} \right)$,
the mutual information $I\left( {{\rm{\bf{x};\bf{r}}}} \right)$ of the quantized DTPC under peak power and average power constraints can be maximized if the input distribution satisfies the following equation %by achieving an input distribution that satisfies the following equation
\begin{align}
\label{equation_17}
{p_{\bf{x}}}\left( x \right) = \frac{{\exp \left( {\eta x} \right){{\prod\nolimits_{\bf{r}} {{P_{{\bf{x}}\left| {\bf{r}} \right.}}\left( {x\left| r \right.} \right)} }^{{g_{{\bf{r}}\left| {\bf{x}} \right.}}\left( {r\left| x \right.} \right)}}}}{{\sum\nolimits_{\bf{x}} {\left( {\exp \left( {\eta x} \right){{\prod\nolimits_{\bf{r}} {{P_{{\bf{x}}\left| {\bf{r}} \right.}}\left( {x\left| r \right.} \right)} }^{{g_{{\bf{r}}\left| {\bf{x}} \right.}}\left( {r\left| x \right.} \right)}}} \right)} }},
\end{align}
where an iterative solution for $\eta $ can be evaluated by the Newton-Raphson method, which is given by
\begin{align}
\label{equation_18} \nonumber
&{\eta ^{\left( k \right)}} = {\eta ^{\left( {k - 1} \right)}}\\
& \!- \!\frac{{\sum\nolimits_{\bf{x}} {\exp \!\left( {{\eta ^{\left( {k - 1} \right)}}x} \right)\!} \left[ {1\! -\! \frac{x}{\epsilon}} \right]\!\prod\nolimits_{\bf{r}} \!{{P_{{\bf{x}}\left| {\bf{r}} \right.}}{{\left( {x\left| r \right.} \right)}^{{g_{{\bf{r}}\left| {\bf{x}} \right.}}\left( {r\left| x \right.} \right)}}} }}{{\sum\nolimits_{\bf{x}} {\!x\!\exp \!\left( {{\eta ^{\left( {k - 1} \right)}}x} \right)\!} \left[ {1 \!- \!\frac{x}{\epsilon}} \right]\!\prod\nolimits_{\bf{r}}\! {{P_{{\bf{x}}\left| {\bf{r}} \right.}}{{\left( {x\left| r \right.} \right)}^{{g_{{\bf{r}}\left| {\bf{x}} \right.}}\left( {r\left| x \right.} \right)}}} }},
\end{align}
where $k$ is the index of iteration.
\begin{proof}
Please see Appendix A.
\end{proof}
\end{corollary}
The Kullback-Leibler (KL) divergence between ${g_{{\bf{r}}\left| {\bf{x}} \right.}}$ and ${g_{\bf{r}}}\left( r \right)$ as well as the gradient descent technique are used to optimize the amplitude ${x_i}$, which is given by
\begin{align}
\label{equation_20} \nonumber
&x_i^{\left( k \right)} = x_i^{\left( {k - 1} \right)} + {\theta ^{\left( {k-1} \right)}} \times \\
&{\left. {\frac{\partial }{{\partial {x_i}}}\!\!\left(\! {\sum\nolimits_{\bf{r}} {{g_{{\bf{r|x}}}}\left( {r{\bf{|}}{x_i}} \right)\log \left( {\frac{{{g_{{\bf{r}}\left| {\bf{x}} \right.}}\left( {r\left| x \right.} \right)}}{{{g_{\bf{r}}}\left( r \right)}}} \right)\! - \!\eta{x_i}} } \!\right)} \right|_{{x_i} = x_i^{\left( {k - 1} \right)}}},
\end{align}
where ${\theta}$ denotes the step-size. ${\theta}$ is selected by the Armijo-type Line Search rule to ensure that the sequence of mutual information values obtained along the iterations is non-decreasing, which ensures convergence to a local maximum.
\begin{proof}
 \emph{Please see Appendix B.}
\end{proof}
By combining (\ref{equation_6}), (\ref{equation_7}), and (\ref{equation_17}), the capacity of the quantized DTPC is obtained as
\begin{equation}
\label{equation_21}
C=\sum\nolimits_{\bf{x}} {\sum\nolimits_{\bf{r}}{p_{\bf{x}}\left( {{x_i}} \right){g_{{\bf{r}}\left| {\bf{x}} \right.}}\left( {r\mid  x_i} \right)} } \log{\frac{{{g_{{\bf{r}}\left| {\bf{x}} \right.}}\!\left( {r\mid x_i} \right)}}{{g_{\bf{r}}\left( r \right)}}}.
\end{equation}
\begin{algorithm}[htbp]
\caption{AO algorithm for the quantized DTPC.}\label{alg:alg1}
\begin{algorithmic}
\STATE
%\STATE {\textsc{TRAIN}}$(\mathbf{X} \mathbf{T})$
\STATE \textbf{Input}: ${\bf{x}} = \left\{ {{x_1},{x_2},...,{x_N}} \right\}$, ${\bf{q}} = \left\{ {{q_1},{q_2},...,{q_{K - 1}}} \right\}$, $A$, $\lambda $
\STATE Initialization (typical values) $s = 1$, $\delta  = {10^{ - 5}}$, $p\left( {{x_1}} \right) = p\left( {{x_2}} \right) = ... = p\left( {{x_N}} \right) = \frac{1}{N}$
\STATE \textbf{repeat}
\STATE \hspace{0.5cm}$k = 1,{\epsilon^{\left( 0 \right)}} = 0$
\STATE \hspace{0.5cm} \textbf{repeat}
\STATE \hspace{1cm}$k = k + 1$
\STATE  \hspace{1cm} \textbf{for} $N_1=0, 1,..., 20$ \textbf{do}
\STATE  \hspace{1.5cm} Compute ${P_{{\bf{x}}\left| {\bf{r}} \right.}}\left( {x\left| r \right.} \right)$ via (\ref{equation_11})
\STATE  \hspace{1.5cm} Compute {$\eta _{}^{\left({k } \right)}$ via (\ref{equation_18})}
\STATE  \hspace{1.5cm} Compute $p_{\bf{x}}^{\left( k \right)}\left( x \right)$ via  (\ref{equation_17})
\STATE  \hspace{1.5cm} Compute $g_{\bf{r}}^{\left( k \right)}\left( r \right)$ via  (\ref{equation_7})
\STATE \hspace{1cm} \textbf{end for}
\STATE \hspace{1cm} \textbf{for} $N_2=0, 1,..., 20$ \textbf{do}
\STATE \hspace{1.5cm} Compute $x_i^{\left( k \right)}$ via (\ref{equation_20})
\STATE \hspace{1cm} \textbf{end for}
\STATE \hspace{1cm} Compute ${\epsilon^{\left( k \right)}} = \sum\nolimits_{\bf{x}} {x_i^{\left( k \right)}{p_{\bf{x}}}\left( {x_i^{\left( k \right)}} \right)}$
\STATE \hspace{0.5cm} \textbf{Until} {$\left| {{\epsilon ^{\left( k \right)}} - {\epsilon ^{\left( {k - 1} \right)}}} \right| \le \delta  \vee k \ge 100$}
\STATE \hspace{0.5cm} Compute $C $ via (\ref{equation_21})
\STATE \hspace{0.5cm} Apply the "two-symbols-one-location strategy" \cite{paper_28}
\STATE \hspace{0.5cm} Re-initialize: $s = \left( {1 - \delta } \right)s$
\STATE \textbf{Until} $s \le {10^{ - 5}}$
\STATE Output: $C$, ${\bf{q}}$, ${\bf{x}}$, ${p_{\bf{x}}}\left( x \right)$
\end{algorithmic}
\label{alg1}
\end{algorithm}

 %for the optimization problem in .
The pseudo-code for the proposed algorithm is detailed in \textbf{Algorithm 1}.
 The proposed algorithm iteratively adjusts parameters such as the input distribution, conditional distribution, and amplitude ${x_i}$ to converge to stable values. In particular, we start with a random initial guess of the input distribution, i.e., $x_i$ and ${p_{\bf{x}}}\left( {{x_i}} \right)$.
For a given variable, i.e., by fixing either $p_{\bf{x}}\left( x \right)$ or ${P_{{\bf{x}}\left| {\bf{r}} \right.}}\left( {x\left| r \right.} \right)$, (\ref{equation_11}) and (\ref{equation_17}) provide methods to alternatively find the optimal solutions for the other variable. Then, (\ref{equation_20}) is used to update the amplitude of the mass point ${x_i}$. The above processes are repeated until the mutual information converges.
Additionally, the mass points may either merge at the same position or diverge depending on the phase transition condition. Thus, a two-symbols-one-location strategy \cite{paper_28} is used to adjust the number of mass points. For example, constellation points within a small distance of each other are merged into common mass points with the sum probability.
 
\begin{figure} [htbp]
\centering
\includegraphics[width=0.45\textwidth]{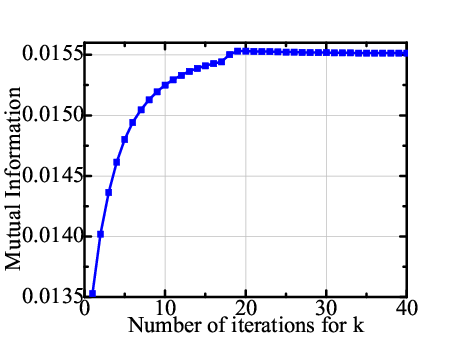}%{eps/capacity_k.eps}
\caption{Capacity for 1-bit quantized DTPC various the number of iterations $k$ in $SNR= 3$ dB.}
 \label{fig_2}
\end{figure}

\textbf{Analysis of convergence.}
Convergence of the gradient descent algorithm to a local optimum is ensured through the proper selection of ${\theta}$ in equation (\ref{equation_20}). The Armijo rule is used to ensure that the sequence of mutual information values obtained along the iterations is non-decreasing, thus ensuring convergence to a local maximum \cite{paper_a01a}.
 For the given variables (input distribution, conditional distribution, and amplitude ${x_i}$), the global optimal solutions for the other variables are obtained. Therefore, the mutual information $I\left( {{\bf{x}};{\bf{r}}} \right)$ increases after each iteration of \textbf{Algorithm 1}.
The results shown in Fig.~\ref{fig_2} are the capaicity of the quantized DTPC for the particular quantization threshold choice $q = 1$.
It shows that $I{\left( {{\bf{x}};{\bf{r}}} \right)^{\left( {k - 1} \right)}} \le I{\left( {{\bf{x}};{\bf{r}}} \right)^{\left( k \right)}},\forall k$, i.e., $I{\left( {{\bf{x}};{\bf{r}}} \right)^{\left( k \right)}}$ is an increasing sequence, and it is upper bounded by a finite number. Thus, $I{\left( {{\bf{x}};{\bf{r}}} \right)^{\left( k \right)}}$ converges to a fixed value when $k \to \infty$.
Additionally, the authors in \cite{paper_25} demonstrated that the BA algorithm converges at a rate inversely proportional to the approximation error. The Newton-Raphson method is utilized to determine $\eta $ in (\ref{equation_17}), which has been shown to converge to a local root provided the initial value is sufficiently close to the desired root \cite{paper_a02a}.
Therefore, the convergence of the proposed algorithm is ensured.

{{\textbf{Complexity Analysis.}
%The proposed AO algorithm exhibits a per-iteration complexity of ${\cal O}\left( {N  K} \right)$, mainly dictated by three key operations: computing conditional probabilities ${P_{{\bf{x}}\left| {\bf{r}} \right.}}\left( {x\left| r \right.} \right)$, updating $\eta$ via Newton-Raphson method, optimizing $p_{\bf{x}}\left( x \right)$, and refining the input symbols $x_i$ through gradient-based techniques. Each step requires traversing all $N$ input symbols and $K$ quantized outputs. 
%Inner loops (21 iterations each) introduce constant factors, while outer-loop termination (geometric decay of $s$) is independent of $N$ and $K$.
%Inner Loops: The fixed 21 iterations per inner loop introduce a constant factor. This does not affect the asymptotic complexity, remaining a multiplier within ${\cal O}$. Outer-Loop Termination (Geometric Decay of $s$): The termination condition is independent of $N$ and $K$, affecting only the total number of iterations, not the per-iteration complexity.
The proposed AO algorithm exhibits a per-iteration complexity of 
${\cal O}\left( {N  K} \right)$, determined by four key operations. First, computing the conditional probability 
${P_{{\bf{x}}\left| {\bf{r}} \right.}}\left( {x\left| r \right.} \right)$ for all $N$ input symbols and $K$ quantized outputs require ${\cal O}\left( {N  K} \right)$ operations. Second, updating the input distribution $p_{\bf{x}}\left( x \right)$ involves exponentiation and normalization, incurring an ${\cal O}\left( {N  K} \right)$ computational complexity. Third, updating the Lagrange multiplier $\eta$ via the Newton-Raphson method requires 
${\cal O}\left( {N } \right)$ operations. Lastly, %refining input amplitudes using gradient-based updates incurs ${\cal O}\left( {K} \right)$ complexity. 
updating the input amplitudes via gradient-based methods entails a computational complexity of ${\cal O}\left( {K} \right)$.
Since the inner loops have a fixed number of iterations, they introduce only a constant factor. Therefore, the overall per-iteration complexity remains 
${\cal O}\left( {N  K} \right)$.
}}
\section{Numerical Results}
{This section presents numerical evaluations to validate the proposed algorithm. 
Specifically, we analyze the capacity-achieving input distributions obtained under 1-bit and 2-bit quantization schemes. To assess performance improvements, the capacity of the quantized DTPC is compared to that of uniform $(K+1)$-ary PAM scheme. In this benchmark scheme, the input distribution comprises equally spaced mass points, each assigned an equal probability.}
\begin{figure} [htbp]
\centering
\includegraphics[width=0.45\textwidth]{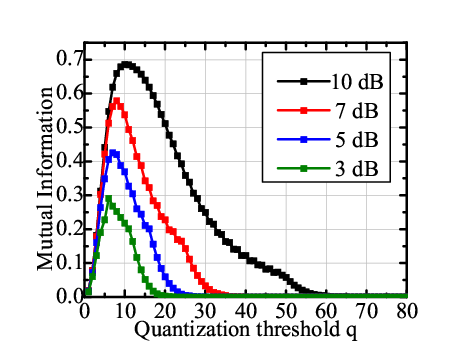}%{eps/quantization_threshold.eps}
\caption{Capacity for 1-bit quantized DTPC various the number of quantization threshold $q$.}
 \label{fig_3}
\end{figure}
%the distributions have mass points with equal probability that are equally spaced.}
%The capacity of the quantized DTPC is  against uniform $(K+1)$-ary PAM schemes, where its distributions have mass points with equal probability that are equally spaced for uniform M-ary PAM.}
\begin{figure*}[htbp]
\centering
\subfloat[]{\includegraphics[width=0.45\textwidth]{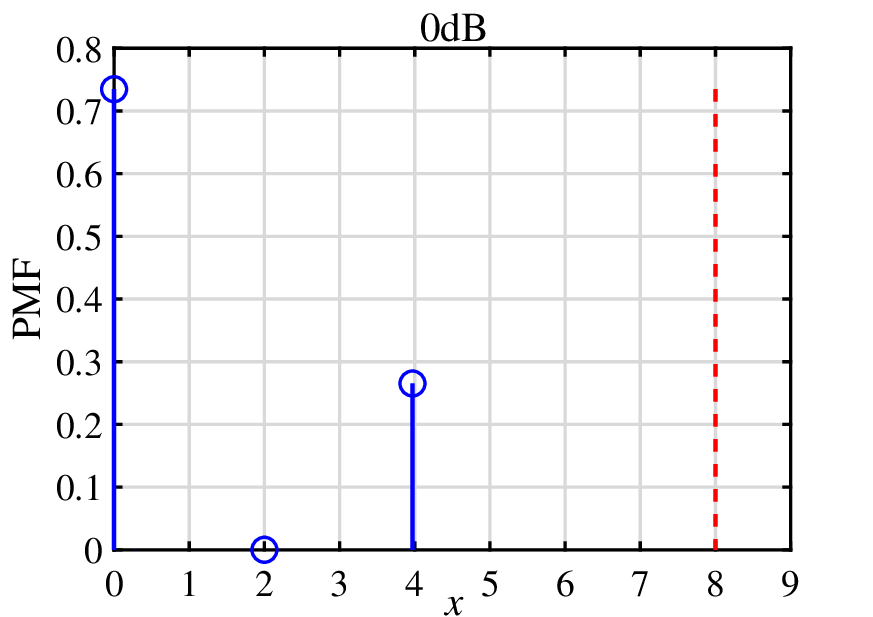}%{eps/fig_2a.eps}
\label{fig_first_case}}
\hfil
\subfloat[]{\includegraphics[width=0.45\textwidth]{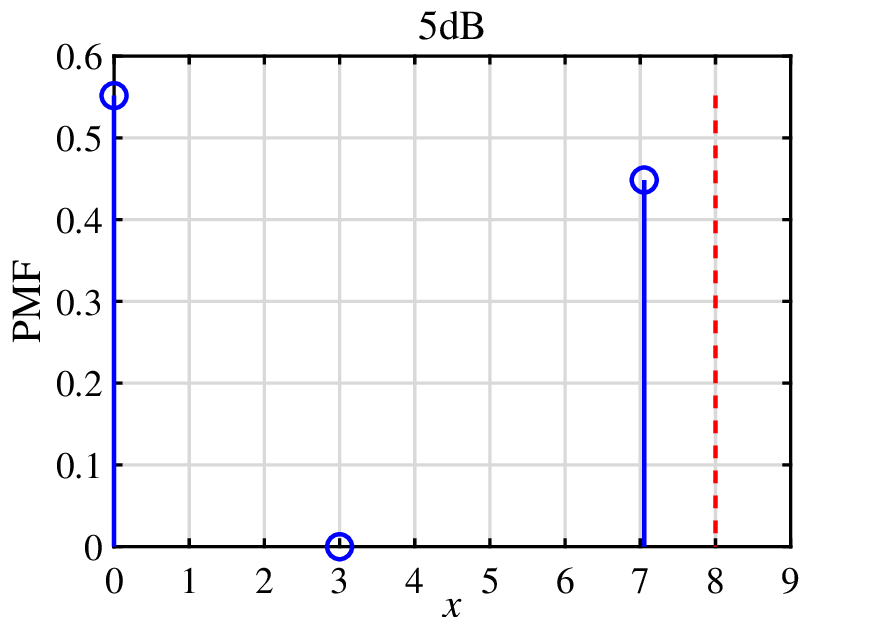}%{eps/1_bit_5db.eps}
\label{fig_second_case}}
\hfil
\subfloat[]{\includegraphics[width=0.45\textwidth]{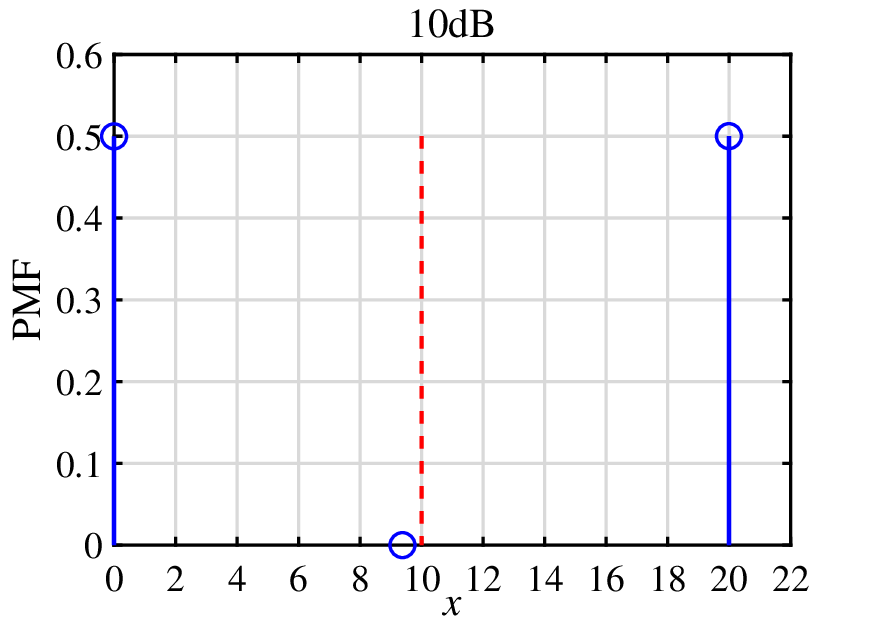}%{eps/fig_2e.eps}
\label{fig_second_case}}
\hfil
\subfloat[]{\includegraphics[width=0.45\textwidth]{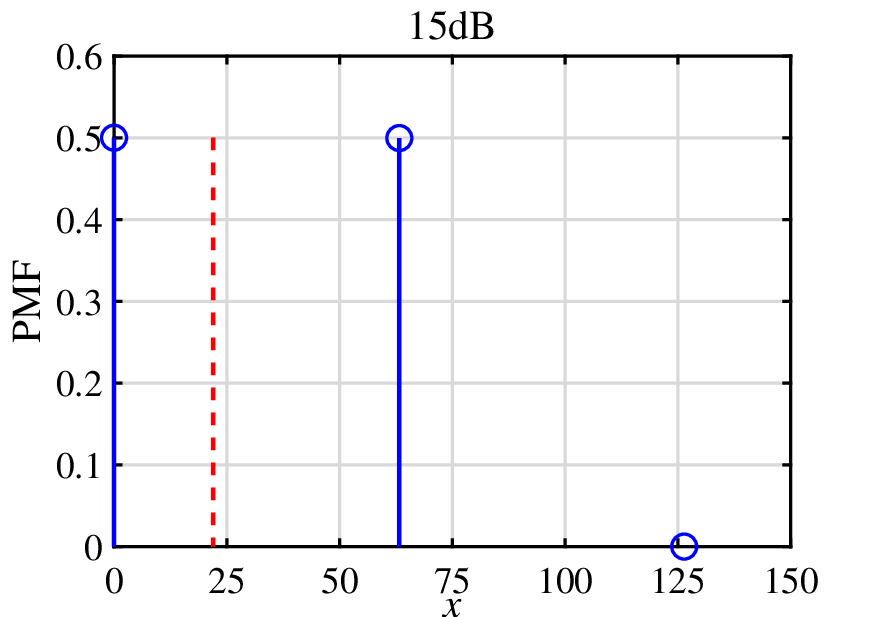}%{eps/fig_2f.eps}
\label{fig_second_case}}
\caption{PMF of the capacity-achieving input distributions with 1-bit ADC quantizer with $\left( {\lambda  = 3,\alpha  = 4} \right)$. (a) $SNR=0$ dB. (b) $SNR=5$ dB. (c) $SNR=10$ dB. (d) $SNR=15$ dB.}
\label{fig_4}
\end{figure*}
\begin{figure*}[htbp]
\centering
\subfloat[]{\includegraphics[width=0.45\textwidth]{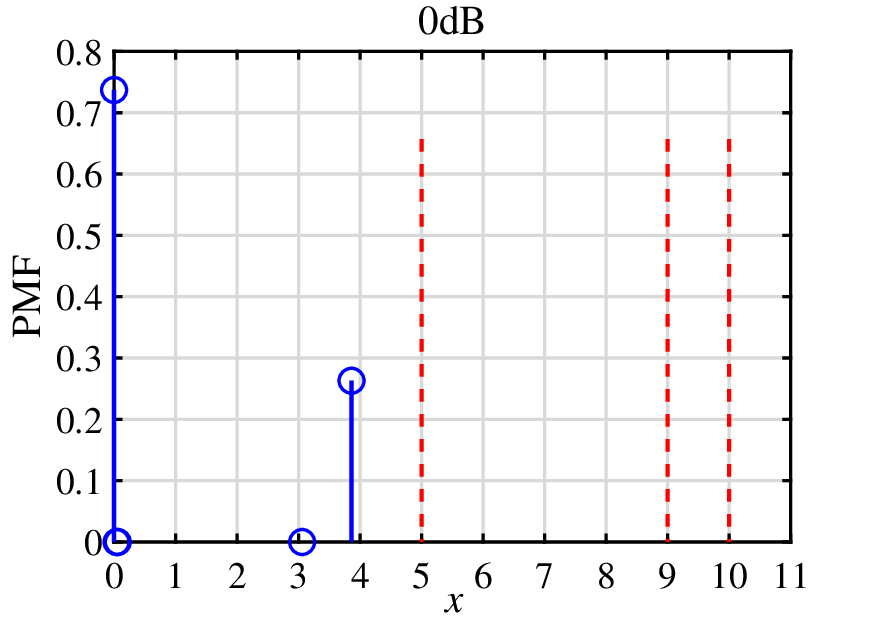}%{eps/fig_3b.eps}
\label{fig_first_case}}
\hfil
\subfloat[]{\includegraphics[width=0.45\textwidth]{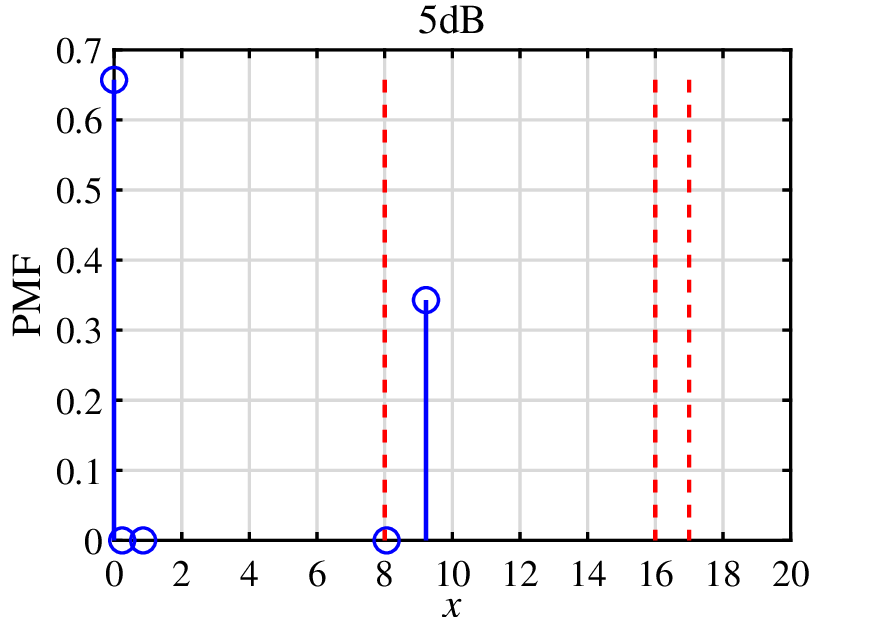}%{eps/2_bit_5db.eps}
\label{fig_second_case}}
\hfil
\subfloat[]{\includegraphics[width=0.45\textwidth]{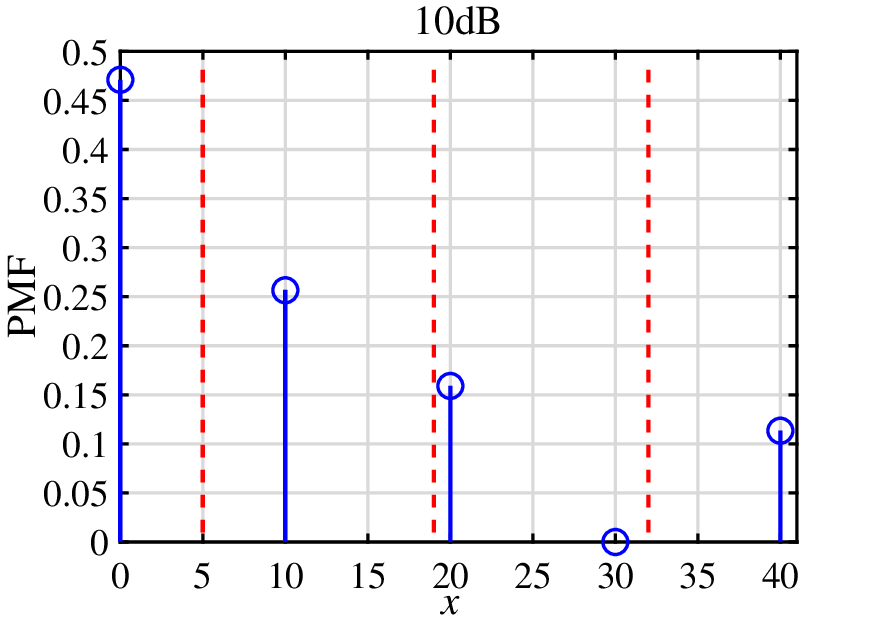}%{eps/fig_3e.eps}
\label{fig_second_case}}
\hfil
\subfloat[]{\includegraphics[width=0.45\textwidth]{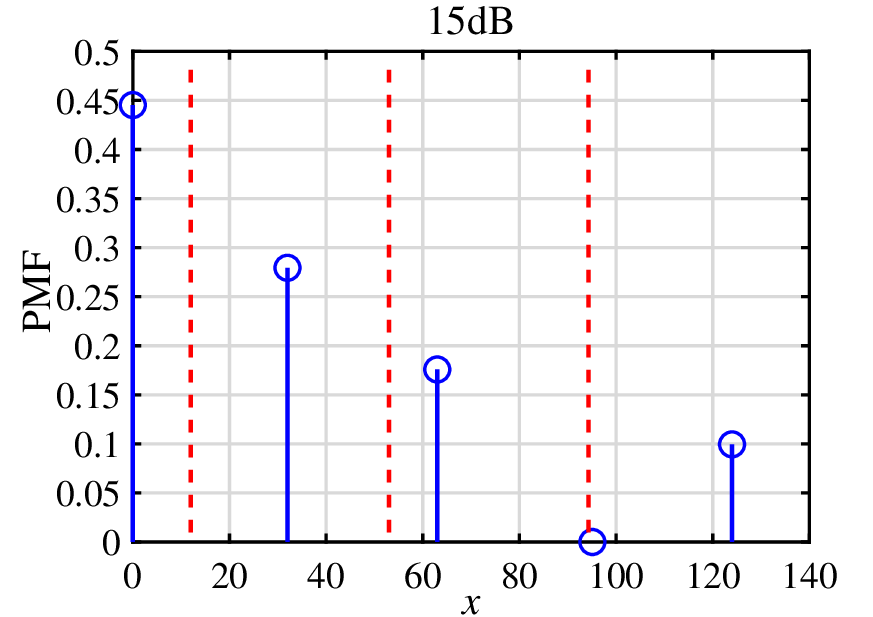}%{eps/fig_3f.eps}
\label{fig_second_case}}
\caption{PMF of the capacity-achieving input distributions with 2-bit ADC quantizer with $\left( {\lambda  = 3,\alpha  = 4} \right)$. (a) $SNR=0$ dB. (b) $SNR=5$ dB. (c) $SNR=10$ dB. (d) $SNR=15$ dB.}
\label{fig_5}
\end{figure*}

Fig.~\ref{fig_3} plots the quantized DTPC capacity as a function of the quantization threshold $q$ at various signal-to-noise ratios (SNRs). An ergodic optimization method is used to find the optimal 1-bit quantizer. It can be observed that for any given SNR, there exists an optimal choice of $q$ that maximizes the channel capacity.
 Furthermore, the variation in channel capacity is minor in the low SNR regime, but becomes more pronounced as the SNR increases, making the selection of the quantizer more critical in the high SNR regime.

The input distribution and quantization thresholds of the 1-bit ADC quantizer at different SNRs are illustrated in Fig.~\ref{fig_4}. In Fig.~\ref{fig_4}, the red lines indicate the quantization thresholds, the blue lines represent the amplitude $x_i$, and the blue dots depict the input distribution. It can be seen that as the SNR increases, the distance between the input mass points and the corresponding thresholds also increases. This leads to an increase in the capacity of the quantized DTPC. It is worth noting that the derived optimal inputs for 1-bit ADC quantized DTPCs under peak power and average power constraints have at most two points. Even if the initial assumption is more relaxed (i.e., assuming the input distribution comprises three mass points), the capacity can be guaranteed to be achievable with at most two points.

Fig.~\ref{fig_5} illustrates the capacity-achieving input distributions for a 2-bit ADC quantizer at different SNRs. As the SNR increases, the distance between the input mass points increases, improving the capacity for the DTPC with 2-bit ADC quantization. Similarly, although the initial input constellation is assumed to consist of five mass points with equal probabilities, the capacity-achieving input distribution of the proposed algorithm has four effective mass points (i.e., non-zero probability). As a result, the proposed algorithm yields a non-uniform input distribution that can guide the design of capacity-achieving input distribution for the quantized DTPC under peak power and average power constraints.
\begin{figure} [htbp]
\centering
\includegraphics[width=0.45\textwidth]{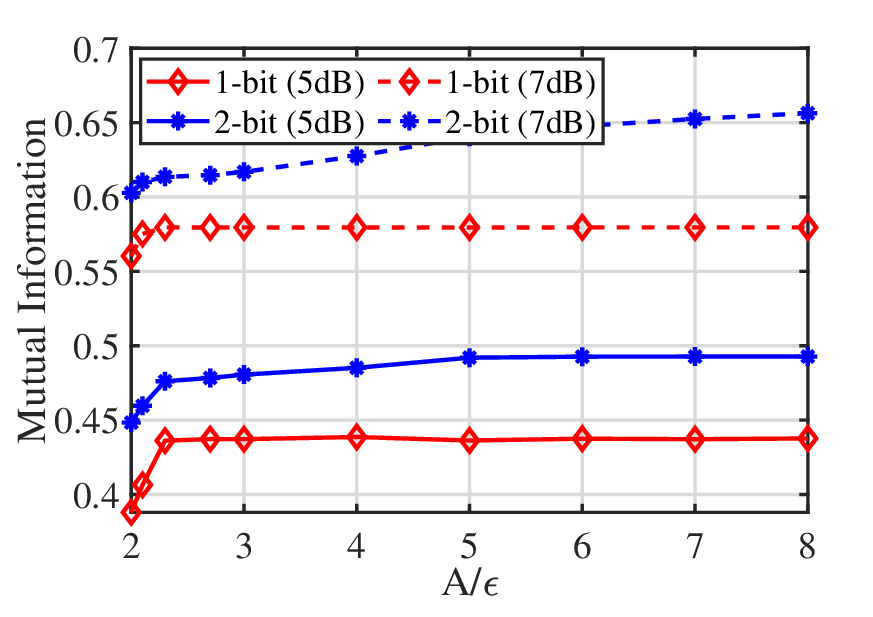}
\caption{Capacity for the 1-bit and 2-bit quantized DTPC with ${\epsilon =5dB}$ and ${\epsilon =7dB}$, respectively.}
 \label{fig_6}
\end{figure}

Fig.~\ref{fig_6} depicts the capacity of low-precision ADC quantized DTPC versus varying peak-to-average power ratios.  It can be observed that the capacities of the 1-bit and 2-bit ADC quantized DTPC
 improve with an increasing ${A \mathord{\left/
 {\vphantom {A \epsilon}} \right.
 \kern-\nulldelimiterspace}\epsilon}$. Therefore, an appropriate increase in peak power enhances capacity with relatively smaller increments in the rate. The capacity converges to a constant value in the high ${A \mathord{\left/
 {\vphantom {A \epsilon}} \right.
 \kern-\nulldelimiterspace}\epsilon}$ region.
 This suggests that beyond a certain point, increasing the peak power of the input signal does not lead to a proportional increase in the capacity of the low-precision quantized DTPC.
 \begin{figure} [htbp]
%\begin{narrow}{-0.125in}{0in}
\centering
\includegraphics[width=0.45\textwidth]{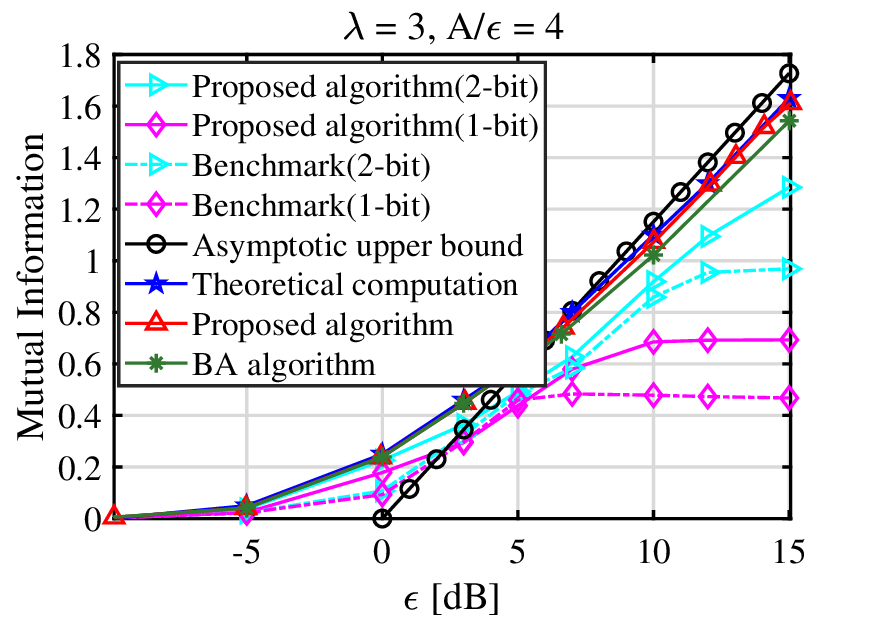}%{eps/Mutual_information.eps}
 \caption{Capacity for the quantized DTPC versus SNRs.}
 \label{fig_7}
%\end{narrow}
\end{figure}

Fig.~\ref{fig_7} shows the capacity of the quantized DTPC versus $\epsilon$ for different quantization precisions. {The results indicate that the proposed algorithm consistently outperforms the benchmark strategies in both 1-bit and 2-bit quantization scenarios, highlighting the advantage of optimizing the input distribution. }It can be observed that the capacities of 1-bit and 2-bit quantized DTPCs are reduced due to quantization errors. Specifically, the capacities of 1-bit ADC and 2-bit ADC achieve $72\%$ and $83\%$ of the theoretical capacity at 5dB, respectively. {Additionally, in the high SNR region (e.g., 12dB), the capacity of the 2-bit ADC quantized DTPC approaches $84\%$ of the theoretical capacity.} This means that high-precision quantizers for the DTPC model can be replaced by low-precision ones with minimal performance loss, thereby significantly reducing hardware costs and power consumption.
For infinite-precision quantized DTPC, the proposed algorithm outperforms the traditional BA algorithm and closely matches the theoretical capacity. Furthermore, the black curve \cite{paper_29a} represents the firm asymptotic upper bounds for the theoretical channel capacity in the high SNR region. It should be noted that the upper bound is valid only in the limit of high SNR for the capacity of the DTPC under average and peak power constraints.
%It can be observed that the capacities of 1-bit and 2-bit quantized DTPCs are reduced due to quantization errors. Specifically, the capacities of 1-bit ADC and 2-bit ADC achieve $72\%$ and $83\%$ of the theoretical capacity at 5dB, respectively. Additionally, in the high SNR region, capacity of the 2-bit ADC quantized DTPC approaches the theoretical capacity, reaching approximately $84\%$ at 12dB. This means that high-precision quantizers for the DTPC model can be replaced by low-precision quantizers with a tolerable performance loss, thereby significantly reducing hardware costs and power consumption.
%For infinite-precision quantized DTPC, the proposed algorithm outperforms the traditional BA algorithm and closely matches the theoretical capacity. Furthermore, the black curve \cite{paper_29a} represents the firm asymptotic upper bounds for the theoretical channel capacity in the high SNR region. It should be noted that the upper bound is valid only in the limit of high SNR for the capacity of the DTPC under average and peak power constraints.
\section{Conclusion}
%This paper considered a low-precision ADC quantized DTPC model under peak power and average power constraints.
This paper investigated a low-precision ADC quantized DTPC model under peak and average power constraints, considering dark current effects.
We formulated the problem of maximizing mutual information over the input distribution and the quantized channel output.
We then proposed an AO algorithm to derive the capacity-achieving input distribution and its optimal amplitude levels. 
%Then, we proposed an AO algorithm to obtain the optimum capacity, capacity-achieving input distribution, and amplitude of the quantized DTPC.
%The AO algorithm based on the Newton-Raphson and the BA methods was proposed to derive the optimum capacity, capacity-achieving input distribution, and amplitude of the quantized DTPC.
Simulation results demonstrated that for the ${\log _2}K$-bit quantized DTPC with dark current, the capacity is achieved through a non-uniform discrete input distribution supported on $K$ mass points under the given power constraints.
%$K$ mass points are required to obtain a non-uniform capacity-achieving input distribution for ${\log _2}K$-bit ADC quantized DTPC with peak power and average power constraints. 
%Simulation results demonstrated that for a log 2K-bit quantized DTPC, the capacity is attained using a non-uniform discrete input distribution supported on K  mass points under the given power constraints.
%The capacity attained by the proposed algorithm closely approaches theoretical values in the case of infinite precision ADC quantization.
The proposed algorithm closely approaches the theoretical capacity in the infinite precision case. 
%The 2-bit ADC achieved about $84\%$ of the theoretical capacity at 12 dB, indicating that low-precision ADCs can effectively replace high-precision ones with minimal performance loss.
%The 2-bit ADC capacity is approximately $84\%$ of the theoretical capacity at 12 dB, which implies that the low-precision quantizers can replace the high-precision ADC quantizers for the DTPC model with a tolerable performance loss.

{\appendix%[Proof of the Zonklar Equations]
\section*{Appendix~A: Proof of equation (\ref{equation_17})} \label{Appendix:A}
\renewcommand{\theequation}{A.\arabic{equation}}
\setcounter{equation}{0}
{The input-quantized output mutual information $I\left( {{\bf{x}};{\bf{r}}} \right)$ is expressed as
{ \begin{align}
\label{equation_A.0}
I\left( {{\bf{x}};{\bf{r}}} \right) = \sum\nolimits_{\bf{x}} {\sum\nolimits_{\bf{r}} {{p_{\bf{x}}}\left( x \right){g_{{\bf{r}}\left| {\bf{x}} \right.}}\left( {r\left| x \right.} \right)\log \frac{{{P_{{\bf{x}}\left| {\bf{r}} \right.}}\left( {x\left| r \right.} \right)}}{{{p_{\bf{x}}}\left( x \right)}}} } .
\end{align} }}
By leveraging \eqref{equation_14} and applying the nonlinear programming theorem, it is given by
%we can derive the following result as
%Based on \eqref{equation_14}, and according to the nonlinear programming theorem, we can obtain as follows
%, $I\left( {{\rm{\bf{x};\bf{r}}}} \right)$ is maximized as
\begin{align}
\label{equation_A.1} \nonumber
&\frac{{\partial W}}{{\partial {p_{\bf{x}}}\left( x \right)}}\!=\! \frac{\partial }{{\partial {p_{\bf{x}}}\left( x \right)}}\!\!\left( \!{\sum\nolimits_{\bf{x}} {\!\sum\nolimits_{\bf{r}} {{p_{\bf{x}}}\left( x \right)\!{g_{{\bf{r}}\left| {\bf{x}} \right.}}\left( {r\left| x \right.} \right)\! \log \! \frac{{{P_{{\bf{x}}\left| {\bf{r}} \right.}}\left( {x\left| r \right.} \right)}}{{{p_{\bf{x}}}\left( x \right)}}} } } \right.\\  \nonumber
&\left. { + {\lambda _0}\sum\nolimits_{\bf{x}} {{p_{\bf{x}}}\left( x \right) - } {\lambda _0} + \eta \sum\nolimits_{\bf{x}} {x{p_{\bf{x}}}\left( x \right)}  - \eta  \epsilon} \right)\\  \nonumber
 &= \!\sum\nolimits_{\bf{r}} {{g_{{\bf{r}}\left| {\bf{x}} \right.}}\left( {r\left| x \right.} \right)\log \frac{{{P_{{\bf{x}}\left| {\bf{r}} \right.}}\left( {x\left| r \right.} \right)}}{{{p_{\bf{x}}}\left( x \right)}}}\\
  &- \sum\nolimits_{\bf{r}} {{p_{\bf{x}}}\left( x \right){g_{{\bf{r}}\left| {\bf{x}} \right.}}\left( {r\left| x \right.} \right)\frac{1}{{{p_{\bf{x}}}\left( x \right)}}}  + {\lambda _0} + \eta x.
\end{align}
{Since the normalization of conditional probability $\sum\nolimits_{\bf{r}} {{g_{{\bf{r}}\left| {\bf{x}} \right.}}\left( {r\left| x \right.} \right)}  =1 $, we can obtain the following 
\begin{align}
\label{A.01}
    \sum\nolimits_{\bf{r}} {{p_{\bf{x}}}\left( x \right){g_{{\bf{r}}\left| {\bf{x}} \right.}}\left( {r\left| x \right.} \right)\frac{1}{{{p_{\bf{x}}}\left( x \right)}}}   = 1.
\end{align}
}
Substituting \eqref{A.01} into \eqref{equation_A.1}, it is given by
%where $\sum\nolimits_{\bf{r}} {{p_{\bf{x}}}\left( x \right){g_{{\bf{r}}\left| {\bf{x}} \right.}}\left( {r\left| x \right.} \right)\frac{1}{{{p_{\bf{x}}}\left( x \right)}}}  = \sum\nolimits_{\bf{r}} {{p_{{\bf{x}},{\bf{r}}}}\left( {x,r} \right)\frac{1}{{{p_{\bf{x}}}\left( x \right)}}}  = 1$ with ${p_{\bf{x}}}\left( x \right) = \sum\nolimits_{\bf{r}} {{p_{{\bf{x}},{\bf{r}}}}\left( {x,r} \right)}$. Therefore, (\ref{equation_A.1}) can be rewritten as
\begin{equation}
\label{equation_A.2}
\frac{{\partial W}}{{\partial {p_{\bf{x}}}\left( x \right)}} = \sum\nolimits_{\bf{r}} {{g_{{\bf{r}}\left| {\bf{x}} \right.}}\left( {r\left| x \right.} \right)\log \frac{{{P_{{\bf{x}}\left| {\bf{r}} \right.}}\left( {x\left| r \right.} \right)}}{{{p_{\bf{x}}}\left( x \right)}}}  - 1 + {\lambda _0} + \eta x.
\end{equation}
Let $\frac{{\partial W}}{{\partial {p_{\bf{x}}}\left( x \right)}} = 0$, (\ref{equation_A.2}) can be calculated as
\begin{equation}
\label{equation_A.3}
\sum\nolimits_{\bf{r}} {{g_{{\bf{r}}\left| {\bf{x}} \right.}}\left( {r\left| x \right.} \right)\log \frac{{{P_{{\bf{x}}\left| {\bf{r}} \right.}}\left( {x\left| r \right.} \right)}}{{{p_{\bf{x}}}\left( x \right)}}}  = 1 - {\lambda _0} - \eta x.
\end{equation}
Thus, (\ref{equation_A.3}) can be calculated as
\begin{align}
\nonumber
&\exp \left( {1 - {\lambda _0} - \eta x} \right)= \exp \left( {\sum\nolimits_{\bf{r}} {{g_{{\bf{r}}\left| {\bf{x}} \right.}}\left( {r\left| x \right.} \right)\log \frac{{{P_{{\bf{x}}\left| {\bf{r}} \right.}}\left( {x\left| r \right.} \right)}}{{{p_{\bf{x}}}\left( x \right)}}} } \right)\\\nonumber
 &= \exp \left( {\sum\nolimits_{\bf{r}} {\log {{\left[ {\frac{{{P_{{\bf{x}}\left| {\bf{r}} \right.}}\left( {x\left| r \right.} \right)}}{{{p_{\bf{x}}}\left( x \right)}}} \right]}^{{g_{{\bf{r}}\left| {\bf{x}} \right.}}\left( {r\left| x \right.} \right)}}} } \right)\\\nonumber
& = {p_{\bf{x}}}{\left( x \right)^{ - \sum\nolimits_{\bf{r}} {{g_{{\bf{r}}\left| {\bf{x}} \right.}}\left( {r\left| x \right.} \right)} }}\prod\nolimits_{\bf{r}} {{P_{{\bf{x}}\left| {\bf{r}} \right.}}{{\left( {x\left| r \right.} \right)}^{{g_{{\bf{r}}\left| {\bf{x}} \right.}}\left( {r\left| x \right.} \right)}}} \\ \label{equation_A.4}
 &= {p_{\bf{x}}}{\left( x \right)^{ - 1}}\prod\nolimits_{\bf{r}} {{P_{{\bf{x}}\left| {\bf{r}} \right.}}{{\left( {x\left| r \right.} \right)}^{{g_{{\bf{r}}\left| {\bf{x}} \right.}}\left( {r\left| x \right.} \right)}}}.
\end{align}
Consequently, the input distribution is obtained by
%equation (\ref{equation_16}) as follows
\begin{equation}
\label{equation_A.5}
{p_{\bf{x}}}\left( x \right) = \frac{{\prod\nolimits_{\bf{r}} {{P_{{\bf{x}}\left| {\bf{r}} \right.}}{{\left( {x\left| r \right.} \right)}^{{g_{{\bf{r}}\left| {\bf{x}} \right.}}\left( {r\left| x \right.} \right)}}} }}{{\exp \left( {1 - {\lambda _0} - \eta x} \right)}}.
\end{equation}
Substituting \eqref{equation_A.5} into \eqref{equation_10} yields the following expressions, which are expressed as
\begin{equation}
\label{equation_B.4}
\sum\nolimits_{\bf{x}} {\frac{{\prod\nolimits_{\bf{r}} {{P_{{\bf{x}}\left| {\bf{r}} \right.}}{{\left( {x\left| r \right.} \right)}^{{g_{{\bf{r}}\left| {\bf{x}} \right.}}\left( {r\left| x \right.} \right)}}} }}{{\exp \left( {1 - {\lambda _0} - \eta x} \right)}}}  = 1,
\end{equation}
\begin{equation}
\label{equation_B.5}
\sum\nolimits_{\bf{x}}{x\frac{{\prod\nolimits_{\bf{r}} {{P_{{\bf{x}}\left| {\bf{r}} \right.}}{{\left( {x\left| r \right.} \right)}^{{g_{{\bf{r}}\left| {\bf{x}} \right.}}\left( {r\left| x \right.} \right)}}} }}{{\exp \left( {1 - {\lambda _0} - \eta x} \right)}}}  \le \epsilon.
\end{equation}
 \eqref{equation_B.4} and \eqref{equation_B.5} can be respectively rewritten as 
{{\begin{equation}
\label{equation_B.6}
{\exp{\left( {1 - {\lambda _0}} \right)}} = \sum\nolimits_{\bf{x}} {\frac{{\prod\nolimits_{\bf{r}} {{P_{{\bf{x}}\left| {\bf{r}} \right.}}{{\left( {x\left| r \right.} \right)}^{{g_{{\bf{r}}\left| {\bf{x}} \right.}}\left( {r\left| x \right.} \right)}}} }}{{\exp \left( { - \eta x} \right)}}},
\end{equation}
}}
\begin{equation}
\label{equation_B.7}
\sum\nolimits_{\bf{x}} {\frac{x}{\epsilon}\frac{{\prod\nolimits_{\bf{r}} {{P_{{\bf{x}}\left| {\bf{r}} \right.}}{{\left( {x\left| r \right.} \right)}^{{g_{{\bf{r}}\left| {\bf{x}} \right.}}\left( {r\left| x \right.} \right)}}} }}{{\exp \left( {1 - {\lambda _0} - \eta x} \right)}}}  \le 1 .
\end{equation}
{Substituting \eqref{equation_B.4} into \eqref{equation_B.7}, it can be derived as
\begin{equation}
\label{equation_B.7b}
\sum\nolimits_{\bf{x}} {\frac{x}{\epsilon}\frac{{\prod\nolimits_{\bf{r}} {{P_{{\bf{x}}\left| {\bf{r}} \right.}}{{\left( {x\left| r \right.} \right)}^{{g_{{\bf{r}}\left| {\bf{x}} \right.}}\left( {r\left| x \right.} \right)}}} }}{{\exp \left( {1 - {\lambda _0} - \eta x} \right)}}}  \le \sum\nolimits_{\bf{x}} {\frac{{\prod\nolimits_{\bf{r}} {{P_{{\bf{x}}\left| {\bf{r}} \right.}}{{\left( {x\left| r \right.} \right)}^{{g_{{\bf{r}}\left| {\bf{x}} \right.}}\left( {r\left| x \right.} \right)}}} }}{{\exp \left( {1 - {\lambda _0} - \eta x} \right)}}}.
\end{equation}
}
%\begin{equation}
%\label{equation_B.7b}
%\sum\nolimits_{\bf{x}} {\frac{x}{\epsilon}\frac{{\prod\nolimits_{\bf{r}} {{P_{{\bf{x}}\left| {\bf{r}} \right.}}{{\left( {x\left| r \right.} \right)}^{{g_{{\bf{r}}\left| {\bf{x}} \right.}}\left( {r\left| x \right.} \right)}}} }}{{\exp \left( { - \eta x} \right)}}}  \le {\exp{\left(1 - {\lambda _0}\right)}}.
%\end{equation}
\eqref{equation_B.7b} can be  rewritten as
\begin{equation}
\label{equation_B.8}
\sum\nolimits_{\bf{x}} {\frac{x}{\epsilon}\frac{{\prod\nolimits_{\bf{r}} {{P_{{\bf{x}}\left| {\bf{r}} \right.}}{{\left( {x\left| r \right.} \right)}^{{g_{{\bf{r}}\left| {\bf{x}} \right.}}\left( {r\left| x \right.} \right)}}} }}{{\exp \left( { - \eta x} \right)}}}  \le \sum\nolimits_{\bf{x}} {\frac{{\prod\nolimits_{\bf{r}} {{P_{{\bf{x}}\left| {\bf{r}} \right.}}{{\left( {x\left| r \right.} \right)}^{{g_{{\bf{r}}\left| {\bf{x}} \right.}}\left( {r\left| x \right.} \right)}}} }}{{\exp \left( { - \eta x} \right)}}} .
\end{equation}
Then, (\ref{equation_B.8}) can be rewritten as
\begin{equation}
\label{equation_B.9}
\sum\nolimits_{\bf{x}} {\exp \left( {\eta x} \right)\left[ {1 - \frac{x}{\epsilon}} \right]\prod\nolimits_{\bf{r}} {{P_{{\bf{x}}\left| {\bf{r}} \right.}}{{\left( {x\left| r \right.} \right)}^{{g_{{\bf{r}}\left| {\bf{x}} \right.}}\left( {r\left| x \right.} \right)}}} }  \ge 0,
\end{equation}
where (\ref{equation_B.9}) is a non-linear equation with the independent variable $\eta$. Define the function $f\left( \eta  \right) = \sum\nolimits_{\bf{x}} {\exp \left( {\eta x} \right)} \left[ {1 - \frac{x}{\epsilon}} \right]\prod\nolimits_{\bf{r}} {{P_{{\bf{x}}\left| {\bf{r}} \right.}}{{\left( {x\left| r \right.} \right)}^{{g_{{\bf{r}}\left| {\bf{x}} \right.}}\left( {r\left| x \right.} \right)}}}$. The Newton-Raphson method is a way to quickly find a good approximation for the root of a real-valued function.
Thus, the iterative solution of $\eta$ is obtained by
\begin{align}
\nonumber \label{equation_B.010}
&{\eta ^{\left( k \right)}} = {\eta ^{\left( {k - 1} \right)}} - \frac{{f\left( {{\eta ^{\left( {k - 1} \right)}}} \right)}}{{f'\left( {{\eta ^{\left( {k - 1} \right)}}} \right)}}\\\nonumber
& = {\eta ^{\left( {k - 1} \right)}} \\ \nonumber
&- \frac{{\sum\nolimits_{\bf{x}} {\exp \left( {{\eta ^{\left( {k - 1} \right)}}x} \right)} \left[ {1 - \frac{x}{\epsilon}} \right]\prod\nolimits_{\bf{r}} {{P_{{\bf{x}}\left| {\bf{r}} \right.}}{{\left( {x\left| r \right.} \right)}^{{g_{{\bf{r}}\left| {\bf{x}} \right.}}\left( {r\left| x \right.} \right)}}} }}{{{{\left[ {\sum\nolimits_{\bf{x}} {\exp \left( {{\eta ^{\left( {k - 1} \right)}}x} \right)} \left[ {1 - \frac{x}{\epsilon}} \right]\prod\nolimits_{\bf{r}} {{P_{{\bf{x}}\left| {\bf{r}} \right.}}{{\left( {x\left| r \right.} \right)}^{{g_{{\bf{r}}\left| {\bf{x}} \right.}}\left( {r\left| x \right.} \right)}}} } \right]}^\prime }}}\\ \nonumber
& = {\eta ^{\left( {k - 1} \right)}}\\ \nonumber
& - \frac{{\sum\nolimits_{\bf{x}} {\exp \left( {{\eta ^{\left( {k - 1} \right)}}x} \right)} \left[ {1 - \frac{x}{\epsilon}} \right]\prod\nolimits_{\bf{r}} {{P_{{\bf{x}}\left| {\bf{r}} \right.}}{{\left( {x\left| r \right.} \right)}^{{g_{{\bf{r}}\left| {\bf{x}} \right.}}\left( {r\left| x \right.} \right)}}} }}{{{{\sum\nolimits_{\bf{x}} {\left[ {\exp \left( {{\eta ^{\left( {k - 1} \right)}}x} \right)} \right]} }^\prime }\left[ {1 - \frac{x}{\epsilon}} \right]\prod\nolimits_{\bf{r}} {{P_{{\bf{x}}\left| {\bf{r}} \right.}}{{\left( {x\left| r \right.} \right)}^{{g_{{\bf{r}}\left| {\bf{x}} \right.}}\left( {r\left| x \right.} \right)}}} }}\\ \nonumber
&= {\eta ^{\left( {k - 1} \right)}} \\ 
&- \frac{{\sum\nolimits_{\bf{x}} {\exp \left( {{\eta ^{\left( {k - 1} \right)}}x} \right)} \left[ {1 - \frac{x}{\epsilon}} \right]\prod\nolimits_{\bf{r}} {{P_{{\bf{x}}\left| {\bf{r}} \right.}}{{\left( {x\left| r \right.} \right)}^{{g_{{\bf{r}}\left| {\bf{x}} \right.}}\left( {r\left| x \right.} \right)}}} }}{{\sum\nolimits_{\bf{x}} {x\exp \left( {{\eta ^{\left( {k - 1} \right)}}x} \right)} \left[ {1 - \frac{x}{\epsilon}} \right]\prod\nolimits_{\bf{r}} {{P_{{\bf{x}}\left| {\bf{r}} \right.}}{{\left( {x\left| r \right.} \right)}^{{g_{{\bf{r}}\left| {\bf{x}} \right.}}\left( {r\left| x \right.} \right)}}} }},
%& = {\eta ^{\left( {k - 1} \right)}} \\ \nonumber
%&- \frac{{\sum\nolimits_{\bf{x}} {\exp \left( {{\eta ^{\left( {k - 1} \right)}}x} \right)} \left[ {1 - \frac{x}{\epsilon}} \right]\prod\nolimits_{\bf{r}} {{P_{{\bf{x}}\left| {\bf{r}} \right.}}{{\left( {x\left| r \right.} \right)}^{{g_{{\bf{r}}\left| {\bf{x}} \right.}}\left( {r\left| x \right.} \right)}}} }}{{{{\sum\nolimits_{\bf{x}} {\left[ {\exp \left( {{\eta ^{\left( {k - 1} \right)}}x} \right)} \right]} }^\prime }\left[ {1 - \frac{x}{\epsilon}} \right]\prod\nolimits_{\bf{r}} {{P_{{\bf{x}}\left| {\bf{r}} \right.}}{{\left( {x\left| r \right.} \right)}^{{g_{{\bf{r}}\left| {\bf{x}} \right.}}\left( {r\left| x \right.} \right)}}} }}\\ \nonumber
%& = {\eta ^{\left( {k - 1} \right)}} \\  \label{equation_B.10}
%&- \frac{{\sum\nolimits_{\bf{x}} {\exp \left( {{\eta ^{\left( {k - 1} \right)}}x} \right)} \left[ {1 - \frac{x}{\epsilon}} \right]\prod\nolimits_{\bf{r}} {{P_{{\bf{x}}\left| {\bf{r}} \right.}}{{\left( {x\left| r \right.} \right)}^{{g_{{\bf{r}}\left| {\bf{x}} \right.}}\left( {r\left| x \right.} \right)}}} }}{{\sum\nolimits_{\bf{x}} {x\exp \left( {{\eta ^{\left( {k - 1} \right)}}x} \right)} \left[ {1 - \frac{x}{\epsilon}} \right]\prod\nolimits_{\bf{r}} {{P_{{\bf{x}}\left| {\bf{r}} \right.}}{{\left( {x\left| r \right.} \right)}^{{g_{{\bf{r}}\left| {\bf{x}} \right.}}\left( {r\left| x \right.} \right)}}} }},
\end{align}
 where $k$ is the index of iteration. 
%\begin{align}
%\label{equation_B.010} \nonumber
%    & {\eta ^{\left( k \right)}}= {\eta ^{\left( {k - 1} \right)}} \\ 
%&- \frac{{\sum\nolimits_{\bf{x}} {\exp \left( {{\eta ^{\left( {k - 1} \right)}}x} \right)} \left[ {1 - \frac{x}{\epsilon}} \right]\prod\nolimits_{\bf{r}} {{P_{{\bf{x}}\left| {\bf{r}} \right.}}{{\left( {x\left| r \right.} \right)}^{{g_{{\bf{r}}\left| {\bf{x}} \right.}}\left( {r\left| x \right.} \right)}}} }}{{{{\sum\nolimits_{\bf{x}} {\left[ {\exp \left( {{\eta ^{\left( {k - 1} \right)}}x} \right)} \right]} }^\prime }\left[ {1 - \frac{x}{\epsilon}} \right]\prod\nolimits_{\bf{r}} {{P_{{\bf{x}}\left| {\bf{r}} \right.}}{{\left( {x\left| r \right.} \right)}^{{g_{{\bf{r}}\left| {\bf{x}} \right.}}\left( {r\left| x \right.} \right)}}} }}\\ \nonumber
%& = {\eta ^{\left( {k - 1} \right)}} \\  
%&- \frac{{\sum\nolimits_{\bf{x}} {\exp \left( {{\eta ^{\left( {k - 1} \right)}}x} \right)} \left[ {1 - \frac{x}{\epsilon}} \right]\prod\nolimits_{\bf{r}} {{P_{{\bf{x}}\left| {\bf{r}} \right.}}{{\left( {x\left| r \right.} \right)}^{{g_{{\bf{r}}\left| {\bf{x}} \right.}}\left( {r\left| x \right.} \right)}}} }}{{\sum\nolimits_{\bf{x}} {x\exp \left( {{\eta ^{\left( {k - 1} \right)}}x} \right)} \left[ {1 - \frac{x}{\epsilon}} \right]\prod\nolimits_{\bf{r}} {{P_{{\bf{x}}\left| {\bf{r}} \right.}}{{\left( {x\left| r \right.} \right)}^{{g_{{\bf{r}}\left| {\bf{x}} \right.}}\left( {r\left| x \right.} \right)}}} }}.
%\end{align}
%Substituting \eqref{equation_B.6} into \eqref{equation_A.5}, 
{Substituting (A.10) into (A.7), the optimum input distribution is given by
%and the optimum input distribution can be obtained by combining \eqref{equation_B.6} in \eqref{equation_A.5}, which is expressed as
\begin{align}    
\label{equation_B.11} \nonumber
&{p_{\bf{x}}}\left( x \right) = \frac{{\prod\nolimits_{\bf{r}} {{P_{{\bf{x}}\left| {\bf{r}} \right.}}{{\left( {x\left| r \right.} \right)}^{{g_{{\bf{r}}\left| {\bf{x}} \right.}}\left( {r\left| x \right.} \right)}}} }}{{\exp \left( {1 - {\lambda _0} - \eta x} \right)}}\\ \nonumber
&{\rm{ = }}\frac{{\exp \left( {\eta x} \right)\prod\nolimits_{\bf{r}} {{P_{{\bf{x}}\left| {\bf{r}} \right.}}{{\left( {x\left| r \right.} \right)}^{{g_{{\bf{r}}\left| {\bf{x}} \right.}}\left( {r\left| x \right.} \right)}}} }}{{\exp \left( {1 - {\lambda _0}} \right)}}\\
&{\rm{ = }}\frac{{\exp \left( {\eta x} \right)\prod\nolimits_{\bf{r}} {{P_{{\bf{x}}\left| {\bf{r}} \right.}}{{\left( {x\left| r \right.} \right)}^{{g_{{\bf{r}}\left| {\bf{x}} \right.}}\left( {r\left| x \right.} \right)}}} }}{{\sum\nolimits_{\bf{x}} {\left( {\exp \left( {\eta x} \right)\prod\nolimits_{\bf{r}} {{P_{{\bf{x}}\left| {\bf{r}} \right.}}{{\left( {x\left| r \right.} \right)}^{{g_{{\bf{r}}\left| {\bf{x}} \right.}}\left( {r\left| x \right.} \right)}}} } \right)} }}.
%&{p_{\bf{x}}}\left( x \right) = \frac{{\exp \left( {\eta x} \right)\prod\nolimits_{\bf{r}} {{P_{{\bf{x}}\left| {\bf{r}} \right.}}{{\left( {x\left| r \right.} \right)}^{{g_{{\bf{r}}\left| {\bf{x}} \right.}}\left( {r\left| x \right.} \right)}}} }}{{\sum\nolimits_{\bf{x}} {\left( {\exp \left( {\eta x} \right)\prod\nolimits_{\bf{r}} {{P_{{\bf{x}}\left| {\bf{r}} \right.}}{{\left( {x\left| r \right.} \right)}^{{g_{{\bf{r}}\left| {\bf{x}} \right.}}\left( {r\left| x \right.} \right)}}} } \right)} }}.
\end{align}}
\begin{proof}
 The proof is completed.
\end{proof}
\numberwithin{equation}{section}
\section*{Appendix~B: Calculation of equation (\ref{equation_20})} \label{Appendix:B}
\renewcommand{\theequation}{B.\arabic{equation}}
\setcounter{equation}{0}
In this appendix, (\ref{equation_20}) can be rewritten as
\begin{align}
\label{equation_C.0} \nonumber
 &x_i^{\left( k \right)}= x_i^{\left( {k - 1} \right)} +\theta ^{\left( {k - 1} \right)}\times \\
&{\left. {\underbrace {\frac{\partial }{{\partial {x_i}}}\left( {\sum\nolimits_{\bf{r}} {{g_{{\bf{r}}|{\bf{x}}}}\left( {r|{x_i}} \right)\log \left( {\frac{{{g_{{\bf{r}}\left| {\bf{x}} \right.}}\left( {r\left| x_i \right.} \right)}}{{{g_{\bf{r}}}\left( r \right)}}} \right) - \eta {x_i}} } \right)}_\Xi } \right|_{{x_i} = x_i^{\left( {k - 1} \right)}}},
\end{align}
{where ${\sum\nolimits_{\bf{r}} {{g_{{\bf{r}}|{\bf{x}}}}\left( {r|{x_i}} \right)\log \left( {\frac{{{g_{{\bf{r}}\left| {\bf{x}} \right.}}\left( {r\left| x_i \right.} \right)}}{{{g_{\bf{r}}}\left( r \right)}}} \right)} }$ is the KL divergence term. It quantifies the discrepancy between the conditional distribution $g_{{\bf{r}}|{\bf{x}}}\left( {r|{x_i}} \right)$ and the marginal distribution ${g_{\bf{r}}}\left( r \right)$. Maximizing this term is equivalent to maximizing the mutual information $I\left( {{\bf{x}};{\bf{r}}} \right)$, which measures the information-theoretic capacity of the communication channel. $- \eta {x_i}$ is a regularization term, which penalizes large amplitudes of ${x_i}$, enforcing a tradeoff between maximizing mutual information and limiting transmit power. The parameter $\eta$ modulates the balance between these competing objectives. 

To guarantee monotonic improvement of the objective function, the step size $\theta ^{\left( {k - 1} \right)}$ is selected using the Armijo line search criterion:
\begin{align}
\label{equation_C.01} \nonumber
&{\theta ^{(k - 1)}} = {\max _{\theta  > 0}}\left\{ {\theta :{ I}\left( {x_i^{(k)}};r \right) \ge { I}\left( {x_i^{(k - 1)};r} \right)} \right.\\
&\left. { + \alpha \theta {{\left\| {\nabla { I}\left( {x_i^{(k - 1)}};r \right)} \right\|}^2}} \right\},
\end{align}
where $\alpha  \in \left( {0,1} \right)$ is a tolerance parameter, and $ {\left\| {\nabla { I}\left( {x_i^{(k - 1)}};r \right)} \right\|}^2$ represents the paradigm square of the gradient. This rule dynamically adjusts $\theta$ to ensure sufficient ascent in mutual information while preventing overshooting.
$\Xi$ in \eqref{equation_C.0} is the gradient term, which can be calculated as}
\begin{align}
\nonumber
&\Xi =\frac{\partial }{{\partial {x_i}}}\left( {\sum\nolimits_r {{g_{{\bf{r|x}}}}\left( {r{\bf{|}}{x_i}} \right)\log {\left( {\frac{{{g_{{\bf{r}}\left| {\bf{x}} \right.}}\left( {r\left| x_i \right.} \right)}}{{{g_{\bf{r}}}\left( r \right)}}} \right)}- \left( {\eta {x_i}} \right)} } \right)\\ \nonumber
%&= \sum\nolimits_r {\frac{\partial }{{\partial {x_i}}}} \left( {{g_{{\bf{r|x}}}}\left( {r{\bf{|}}{x_i}} \right)\log \left( {\frac{{{p_{\bf{x}}}\left( {{x_i}} \right)}}{{{g_{\bf{r}}}\left( r \right)}}{g_{{\bf{r|x}}}}\left( {r{\bf{|}}{x_i}} \right)} \right)} \right) - \eta\\ \nonumber
 &= \sum\nolimits_{\bf{r}} {\left[ {\underbrace {\frac{{\partial {g_{{\bf{r}}|{\bf{x}}}}\left( {r|{x_i}} \right)}}{{\partial {x_i}}}}_{{\Xi _1}}\log {\left( {\frac{{{g_{{\bf{r}}\left| {\bf{x}} \right.}}\left( {r\left| x_i \right.} \right)}}{{{g_{\bf{r}}}\left( r \right)}}} \right)}} \right.} \\ \label{equation_C.1}
 &+ \left. {\underbrace {\frac{{\partial \log {\left( {\frac{{{g_{{\bf{r}}\left| {\bf{x}} \right.}}\left( {r\left| x_i \right.} \right)}}{{{g_{\bf{r}}}\left( r \right)}}} \right)}}}{{\partial {x_i}}}}_{{\Xi _2}}{g_{{\bf{r}}|{\bf{x}}}}\left( {r|{x_i}} \right)} \right] - \eta,
\end{align}
where ${\Xi _1}$ in (\ref{equation_C.1}) is obtained by
{\begin{align}
\label{equation_C.2} \nonumber
&{\Xi _1}=\frac{{\partial {g_{{\bf{r|x}}}}\left( {r{\bf{|}}{x_i}} \right)}}{{\partial {x_i}}} = \sum\limits_{{y_j} = {q_{i - 1}}}^{{q_i}} {\frac{{\partial \left( {\frac{{{{\left( {x + \lambda } \right)}^{{y_j}}}{\exp \left( { - (x + \lambda )} \right)}}}{{{y_j}!}}} \right)}}{{\partial {x_i}}}}\\ \nonumber\nonumber
&= \sum\limits_{{y_j} = {q_{i - 1}}}^{{q_i}} \exp \left( { - (x + \lambda )} \right){\left( {\frac{{{y_j}{{\left( {x + \lambda } \right)}^{{y_j} - 1}}}}{{{y_j}!}} - \frac{{{{\left( {x + \lambda } \right)}^{{y_j}}}}}{{{y_j}!}}{}} \right)} \\ \nonumber
 &= \sum\limits_{{y_j} = {q_{i - 1}}}^{{q_i}} {\exp } \left( { - (x + \lambda )} \right)\frac{{{{\left( {x + \lambda } \right)}^{{y_j}}}}}{{{y_j}!}}\left( {\frac{{{y_j}}}{{\left( {x + \lambda } \right)}} - 1} \right)\\
&{\rm{ = }}\sum\limits_{{y_j} = {q_{i - 1}}}^{{q_i}} {{P_{{\rm{y}}\mid {\bf{x}}}}(y\mid {x_i})} \left( {\frac{{{y_j}}}{{\left( {x + \lambda } \right)}} - 1} \right).
\end{align}}
${\Xi _2}$ in (\ref{equation_C.1}) is given by
%\begin{align}
%\nonumber
%&{\Xi _2}=\frac{{\partial \log {\left( {\frac{{{g_{{\bf{r}}\left| {\bf{x}} \right.}}\left( {r\left| x \right.} \right)}}{{{g_{\bf{r}}}\left( r \right)}}} \right)}}}{{\partial {x_i}}}\\ \nonumber
%& =\frac{{{g_{\bf{r}}}\left( r \right)}}{{{p_{\bf{x}}}\left( {{x_i}} \right){g_{{\bf{r|x}}}}\left( {r{\bf{|}}{x_i}} \right)}} \times \frac{{\partial \left( {\frac{{{p_{\bf{x}}}\left( {{x_i}} \right)}}{{{g_{\bf{r}}}\left( r \right)}}{g_{{\bf{r|x}}}}\left( {r{\bf{|}}{x_i}} \right)} \right)}}{{\partial {x_i}}}\\ \nonumber
 %&= \sum\limits_{{y_j} = {q_{i - 1}}}^{{q_i}} {\left( {\frac{{\left( {{y_j} - {x - \lambda } } \right){{\left( {x + \lambda } \right)}^{{y_j} - 1}}}}{{{y_j}!}}} \right)} {\exp \left( { - (x + \lambda )} \right)}\\ \label{equation_C.3}
 %&\times{\left( {\sum\limits_{{y_j} = {q_{i - 1}}}^{{q_i}} {\frac{{{{\left( {x + \lambda } \right)}^{{y_j}}}{\exp \left( { - (x + \lambda )} \right)}}}{{{y_j}!}}} } \right)^{{\rm{ - }}1}}.
%\end{align}
\begin{align}
 \label{equation_C.3} \nonumber
&{\Xi _2} = \frac{{\partial \log \left( {\frac{{{g_{{\bf{r}}\left| {\bf{x}} \right.}}\left( {r\left| x_i \right.} \right)}}{{{g_{\bf{r}}}\left( r \right)}}} \right)}}{{\partial {x_i}}}= \frac{{{g_{\bf{r}}}\left( r \right)}}{{{g_{{\bf{r}}|{\bf{x}}}}\left( {r|x_i} \right)}} \times \frac{{\partial \left( {\frac{{{g_{{\bf{r}}\left| {\bf{x}} \right.}}\left( {r\left| x_i \right.} \right)}}{{{g_{\bf{r}}}\left( r \right)}}} \right)}}{{\partial {x_i}}}\\
 &= \frac{1}{{{g_{{\bf{r}}|{\bf{x}}}}\left( {r|x_i} \right)}}{\Xi _1}.
\end{align}
 Consequently, (\ref{equation_C.1}) can be rewritten as
 {\begin{align}
\nonumber \label{equation_C.4}
%&\frac{\partial }{{\partial {x_i}}}\left( {\sum\nolimits_{\bf{r}} {{g_{{\bf{r}}|{\bf{x}}}}\left( {r|{x_i}} \right)\log \left( {\frac{{{g_{{\bf{r}}|{\bf{x}}}}\left( {r|{x_i}} \right)}}{{{g_{\bf{r}}}\left( r \right)}}} \right) - \left( {\eta {x_i}} \right)} } \right)\\
&\Xi  = \sum\nolimits_{\bf{r}} {\left[ {\sum\limits_{{y_j} = {q_{i - 1}}}^{{q_i}} {{P_{{\rm{y}}\mid {\bf{x}}}}(y\mid {x_i})\left( {\frac{{{y_j}}}{{\left( {{x_i} + \lambda } \right)}} - 1} \right)} } \right.} \\
 &\times \left. {\left( {\log \left( {\frac{{{g_{{\bf{r}}|{\bf{x}}}}\left( {r|{x_i}} \right)}}{{{g_{\bf{r}}}\left( r \right)}}} \right) + 1} \right)} \right] - \eta  .
 \end{align}}
Thus, (\ref{equation_20}) can be rewritten as
{\begin{align}
\nonumber \label{equation_C.5}
&x_i^{\left( k \right)} = x_i^{\left( {k - 1} \right)} + {\theta ^{\left( {k - 1} \right)}}\left[ {\sum\nolimits_{\bf{r}} {\left[ {\sum\limits_{{y_j} = {q_{i - 1}}}^{{q_i}} {{P_{{\rm{y}}\mid {\bf{x}}}}(y\mid {x_i})} } \right.} } \right.  \\ 
&\left. { \times \left( {\frac{{{y_j}}}{{\left( {{x_i} + \lambda } \right)}} - 1} \right)\!\left. {\left( {\log \left( {\frac{{{g_{{\bf{r}}|{\bf{x}}}}\left( {r|{x_i}} \right)}}{{{g_{\bf{r}}}\left( r \right)}}} \right) + 1} \right)} \right] - \eta } \right].
\end{align}}
\section{References Section}
%\section{Simple References}
%You can manually copy in the resultant .bbl file and set second argument of $\backslash${\tt{begin}} to the number of references
 %(used to reserve space for the reference number labels box).
%\balance

}

% your own custom macro containing the $\backslash${\tt{includegraphics}} command to make things
%\bf{If you include a photo:}\vspace{-33pt}
%\begin{IEEEbiography}[{\includegraphics[width=1in,height=1.25in,clip,keepaspectratio]{fig1}}]{Michael Shell}
%Use $\backslash${\tt{begin\{IEEEbiography\}}} and then for the 1st argument use $\backslash${\tt{includegraphics}} to declare and link the author photo.
%Use the author name as the 3rd argument followed by the biography text.
%\end{IEEEbiography}

\vfill

\end{document}